\documentclass[reprint,amssymb, amsmath, aps, superscriptaddress, showpacs, footinbib,  prb]{revtex4-1}
\pdfoutput=1
\usepackage{graphicx}

\newcommand{\eff}{\text{eff}}

\newcommand{\iso}{\text{iso}}

\newcommand{\Cels}{$^{\circ}$C}

\begin{document}

\title{Frustrated pentagonal Cairo lattice in the non-collinear antiferromagnet Bi$_4$Fe$_5$O$_{13}$F}

\author{Artem M. Abakumov}
\email{Artem.Abakumov@ua.ac.be}

\author{Dmitry Batuk}
\affiliation{EMAT, University of Antwerp, Groenenborgerlaan 171, B-2020 Antwerp, Belgium}

\author{Alexander A. Tsirlin}
\email{altsirlin@gmail.com}
\affiliation{Max Planck Institute for Chemical Physics of Solids, N\"{o}thnitzer
Str. 40, 01187 Dresden, Germany}
\affiliation{National Institute of Chemical Physics and Biophysics, 12618 Tallinn, Estonia}

\author{Clemens Prescher}
\author{Leonid Dubrovinsky}
\affiliation{Bayerisches Geoinstitut, Universit\"{a}t Bayreuth, 95440 Bayreuth, Germany}

\author{Denis V. Sheptyakov}
\affiliation{Laboratory for Neutron Scattering, Paul Scherrer Institut, CH-5232 Villigen, Switzerland}

\author{Walter Schnelle}
\affiliation{Max Planck Institute for Chemical Physics of Solids, N\"{o}thnitzer
Str. 40, 01187 Dresden, Germany}

\author{Joke Hadermann}
\author{Gustaaf Van Tendeloo}
\affiliation{EMAT, University of Antwerp, Groenenborgerlaan 171, B-2020 Antwerp, Belgium}


\begin{abstract}
The crystal and magnetic structures and underlying magnetic interactions of Bi$_4$Fe$_5$O$_{13}$F, a model system for studying the physics of the Cairo pentagonal spin lattice, are investigated by transmission electron microscopy, low-temperature synchrotron x-ray and neutron powder diffraction, thermodynamic measurements, and density functional band-structure calculations. The crystal structure of Bi$_4$Fe$_5$O$_{13}$F contains infinite rutile-like chains of edge-sharing FeO$_6$ octahedra interconnected by the Fe$_2$O$_7$ groups of two corner-sharing FeO$_4$ tetrahedra. The cavities between the chains are filled with the fluorine-centered Bi$_4$F tetrahedra. The Fe$^{3+}$ cations form pentagonal units that give rise to an unusual topology of frustrated exchange couplings and underlie a sequence of the magnetic transitions at $T_1$ = 62~K, $T_2$ = 71~K, and $T_N$ = 178~K. Below $T_1$, Bi$_4$Fe$_5$O$_{13}$F forms a fully ordered non-collinear antiferromagnetic structure, whereas the magnetic state between $T_1$ and $T_N$ may be partially disordered according to the sizable increase in the magnetic entropy at $T_1$ and $T_2$. Therefore, Bi$_4$Fe$_5$O$_{13}$F shows the evidence of intricate magnetic transitions that were never anticipated for the pentagonal Cairo spin lattice. Additionally, it manifests a sillimanite (Al$_2$SiO$_5$)-based homologous series of compounds that feature the pentagonal magnetic lattice spaced by a variable number of octahedral units along the rutile-type chains.
\end{abstract}

\maketitle

\section{Introduction}
\label{introduction}
Magnetically frustrated systems entail a pattern of interactions that can not be satisfied simultaneously.\cite{moessner2006,balents2010} These patterns typically involve equivalent antiferromagnetic (AFM) interactions arranged in closed loops with an odd number of bonds, as in a pentagon or in a triangle. As one moves around the loop, the spin flips on each bond, and the translation symmetry is violated because the full turn around the loop leads to the inverted spin direction. Therefore, in frustrated systems spins can not develop a simple collinear AFM order, instead more complex magnetic ground states appear.\cite{balents2010} Finding appropriate model systems for experimental studies of frustrated magnetism is a long-standing challenge.\cite{ramirez1994,greedan2001,harrison2004,gardner2010} The availability of model systems is crucial for addressing theoretical problems and raising new questions to theoretical as well as experimental research.

The loops with an odd number of bonds are typically triangular. Planar triangular spin lattices have been widely studied \cite{moessner2006,balents2010,zhou2011,cheng2011} in at attempt to achieve a spin-liquid state that lacks long-range magnetic order down to zero temperature. When triangles form corner-sharing tetrahedra instead of tiling the plane, the system enters a so-called spin-ice state that facilitates experimental access to magnetic monopoles.\cite{moessner2006,balents2010,gardner2010,castelnovol2008} Thanks to a different topology, spin lattices based on pentagons could lead to a range of hitherto unexplored properties and ground states. However, experimental and even theoretical considerations of such lattices remain challenging because the five-fold symmetry of an ideal pentagon excludes the periodicity of the lattice.

\begin{figure}
\includegraphics{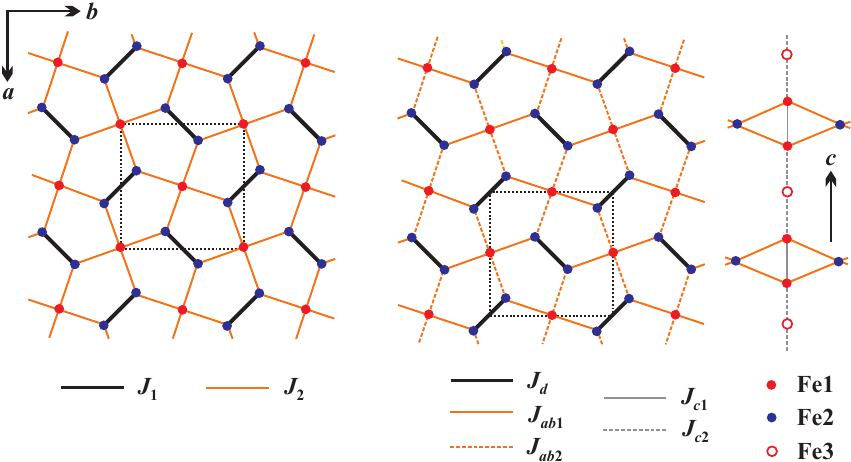}
\caption{\label{lattice}
(Color online) Cairo pentagonal lattice (left panel) and its derivative version in Bi$_4$Fe$_5$O$_{13}$F (right panel). Note that the ideal lattice comprises only two couplings, $J_1$ and $J_2$, whereas in Bi$_4$Fe$_5$O$_{13}$F $J_2$ splits into $J_{ab1}$ and $J_{ab2}$. Additionally, the couplings $J_{c1}$ and $J_{c2}$ along the $c$ direction are present.
}
\end{figure}

Pentagonal units do not form a regular tiling in the plane. To build a tiling, the pentagons have to be distorted and should, unlike the triangles, involve non-equivalent bonds. One of the least-perimeter tilings allowed for pentagons is known as the Cairo lattice. It entails the coupling $J_1$ between three-vertex sites, and the coupling $J_2$ running from each three-vertex site to the surrounding four-vertex sites (Fig.~\ref{lattice}). Recent theoretical work puts forward a number of unconventional ground states, including spin nematics, and other interesting properties of the Cairo spin lattice.\cite{ralko2011,rousochatzakis2012} These predictions remain highly challenging for experimental verification. In crystals, the pentagonal arrangement of magnetic ions is very rare, which is partly related to the fact that periodic crystals can not have five-fold symmetry. Presently, the only known material prototype of the Cairo lattice is Bi$_2$Fe$_4$O$_9$ having mullite-type crystal structure.\cite{niizeki1968,tutov1970,ressouche2009}

While theory readily spans the broad range of the $J_1$ and $J_2$ parameters,\cite{ralko2011,rousochatzakis2012}  Bi$_2$Fe$_4$O$_9$ represents a single point on the effective $J_1$-$J_2$ diagram.\cite{ressouche2009} Experimental access to the diverse magnetism of the Cairo lattice requires a range of compounds with different $J_1$ and $J_2$. In the present paper, we demonstrate that a deliberate modification of the sillimanite-type (Al$_2$SiO$_5$) structure may lead to interesting materials showing exotic magnetism of the Cairo spin lattice. Specifically, we report the crystal structure and magnetic behavior of the hitherto unknown oxyfluoride Bi$_4$Fe$_5$O$_{13}$F that undergoes a sequence of intricate magnetic transitions and reveals unconventional magnetism pertaining to the pentagonal arrangement of magnetic ions in a crystal. 
\begin{table*}
\begin{minipage}{16cm}
\caption{\label{rietveld parameters}
Selected parameters of the Rietveld refinements of the Bi$_4$Fe$_5$O$_{13}$F structure from the NPD data.}
\begin{ruledtabular}
\begin{tabular}{cccc}
	Formula & & Bi$_4$Fe$_5$O$_{13}$F & \\
	Space group & & $P4_2/mbc$ & \\
  $T$ (K) & 1.5 & 300 & 700 \\
  $a$ (\r A) & 8.28671(9) & 8.2995(1) & 8.3246(1) \\
  $b$ (\r A) & 18.0501(2) & 18.0573(3) & 18.1173(3) \\
  $Z$ & & 4 & \\
  Cell volume (\r A$^3$) & 1239.49(2) & 1243.81(3) & 1255.52(4) \\
  Calculated density (g/cm$^3$) & 7.188 & 7.165 & 7.098 \\
  Radiation & & Neutron, $\lambda=1.8857$~\r A & \\
  $R_F$, $R_P$, $R_{wP}$ & 0.013, 0.030, 0.040 & 0.013, 0.029, 0.037 & 0.018, 0.033, 0.042 \\
\end{tabular}
\end{ruledtabular}
\end{minipage}
\end{table*}
 
\section{Methods}
\label{methods}
Powder samples of Bi$_4$Fe$_5$O$_{13}$F were prepared by a solid-state reaction of Bi$_2$O$_3$ (99.9~\% Aldrich), Fe$_2$O$_3$ (nanopowder $<$50~nm, Aldrich) and BiF$_3$ (99.99~\% Aldrich). The initial compounds were weighed in stoichiometric ratios, mixed, rigorously ground and pressed into pellets. The pellets were placed into alumina crucibles, covered with a lid, and then sealed in quartz tubes under a dynamical vacuum of ~10$^{-3}$~mbar. The samples were annealed at 650~\Cels\ for 3~h and at 750~\Cels\ for 6~h with intermediate regrinding.

X-ray powder diffraction (XRPD) measurements were conducted with a Huber G670 Guinier diffractometer (CuK$_{\alpha1}$ radiation, curved Ge(111) monochromator, transmission mode, image plate). High-resolution synchrotron X-ray powder diffraction (SXPD) data were collected at the ID31 beamline of European Synchrotron Radiation Facility (ESRF, Grenoble, France) using a wavelength of 0.40006~\r A and eight scintillation detectors, each preceded by a Si(111) analyzer crystal. The powder sample was contained in a thin-walled quartz capillary that was spun during the experiment. The temperature of the sample was controlled by a He-flow cryostat (temperature range 10 - 300~K) and hot-air blower (300 - 700~K).

Neutron powder diffraction (NPD) data were collected with the high-resolution powder diffractometer HRPT (Paul Scherrer Institut (PSI), Switzerland) at the wavelength of 1.8857~\r A with the use of a standard orange cryostat and radiation-type furnace to cover the temperature range from 1.5 to 700~K. Structure solution was performed with direct methods using the EXPO software.\cite{altomare1999} The Rietveld refinement of the crystal and magnetic structure against the powder diffraction data was performed with the JANA2006 package.\cite{jana2006}

Transmission electron microscopy (TEM) investigation was performed with a FEI Tecnai G$^2$ microscope operated at 200~kV. The TEM specimen was prepared by grinding the sample in ethanol and depositing the dispersion on a copper grid covered with a holey carbon film. The sample was analyzed by means of electron diffraction (ED) and high resolution high angle annular dark field scanning transmission electron microscopy (HAADF-STEM). The simulated HAADF-STEM images were calculated using the QSTEM 2.0 software.\cite{koch2002} The Bi and Fe content has been measured by energy dispersive X-ray (EDX) analysis performed with a Jeol JEM5510 scanning electron microscope equipped with an INCA EDX system (Oxford instruments). The analysis has been done using the Bi-M and Fe-K lines. More than 50 spectra have been collected in order to obtain a statistically relevant result. Within the experimental error, the measured Bi/Fe = 0.84(6) ratio agrees with the nominal chemical composition.

The magnetic susceptibility of Bi$_4$Fe$_5$O$_{13}$F was measured with a Quantum Design MPMS SQUID magnetometer in the temperature range $2-600$~K in magnetic fields up to 5~T. Heat capacity was measured by a relaxation technique using a Quantum Design PPMS in the temperature range $1.8-320$~K.

The M\"{o}ssbauer spectrum was recorded at room temperature (293~K) in transmission mode using a constant-acceleration M\"{o}ssbauer spectrometer with a nominal 2.02 GBq $^{57}$Co source in a 6~$\mu$m Rh matrix (conventional source). The velocity scale was calibrated relative to a 30~$\mu$m $\alpha$-Fe foil using the positions certified for the National Bureau of Standards standard reference ma-terial no.~1541; a line width of 0.30 mm/s for the outer lines of $\alpha$-Fe was obtained at room temperature. The spectrum collection time was 10 days. The M\"{o}ssbauer spectrum was fitted using the MossA software package.\cite{prescher2012}

The electronic structure of Bi$_4$Fe$_5$O$_{13}$F was calculated within the density functional theory (DFT) framework using the FPLO code\cite{fplo} and the Perdew-Wang parametrization of the exchange-correlation potential.\cite{pw92} The symmetry-irreducible part of the first Brillouin zone was sampled with a $k$ mesh of 72 points for the unit cell with 92 atoms. Strong correlations in the Fe $3d$ shell were treated by the DFT+$U$ procedure with a Coulomb repulsion $U=7$~eV and Hund's exchange $J=1$~eV.\cite{spiel2009} Magnetic couplings were evaluated by mapping total energies of collinear spin configurations onto the Heisenberg model.

\section{Results}
\label{results}
\subsection{Crystal structure}

\begin{figure}
\includegraphics{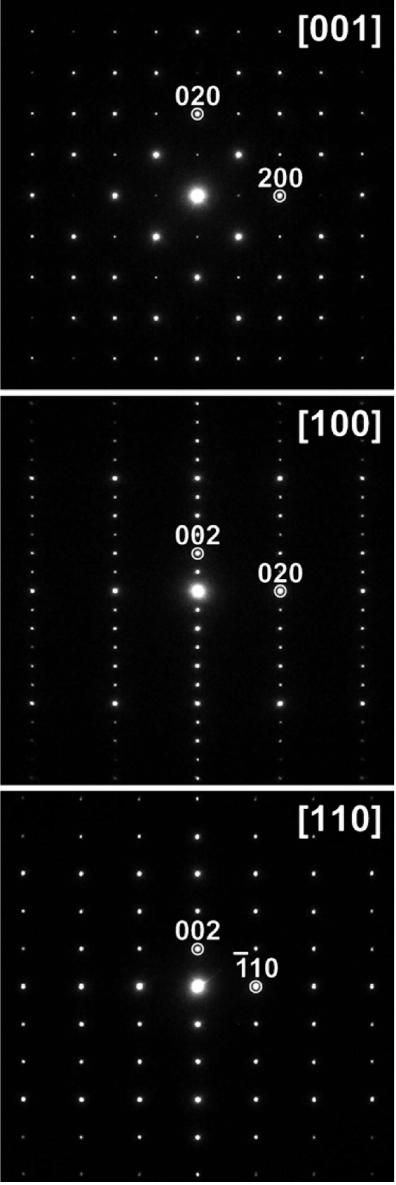}
\caption{\label{electron diffraction}
ED patterns of Bi$_4$Fe$_5$O$_{13}$F along main zone axes.
}
\end{figure}

\begin{figure}
\includegraphics{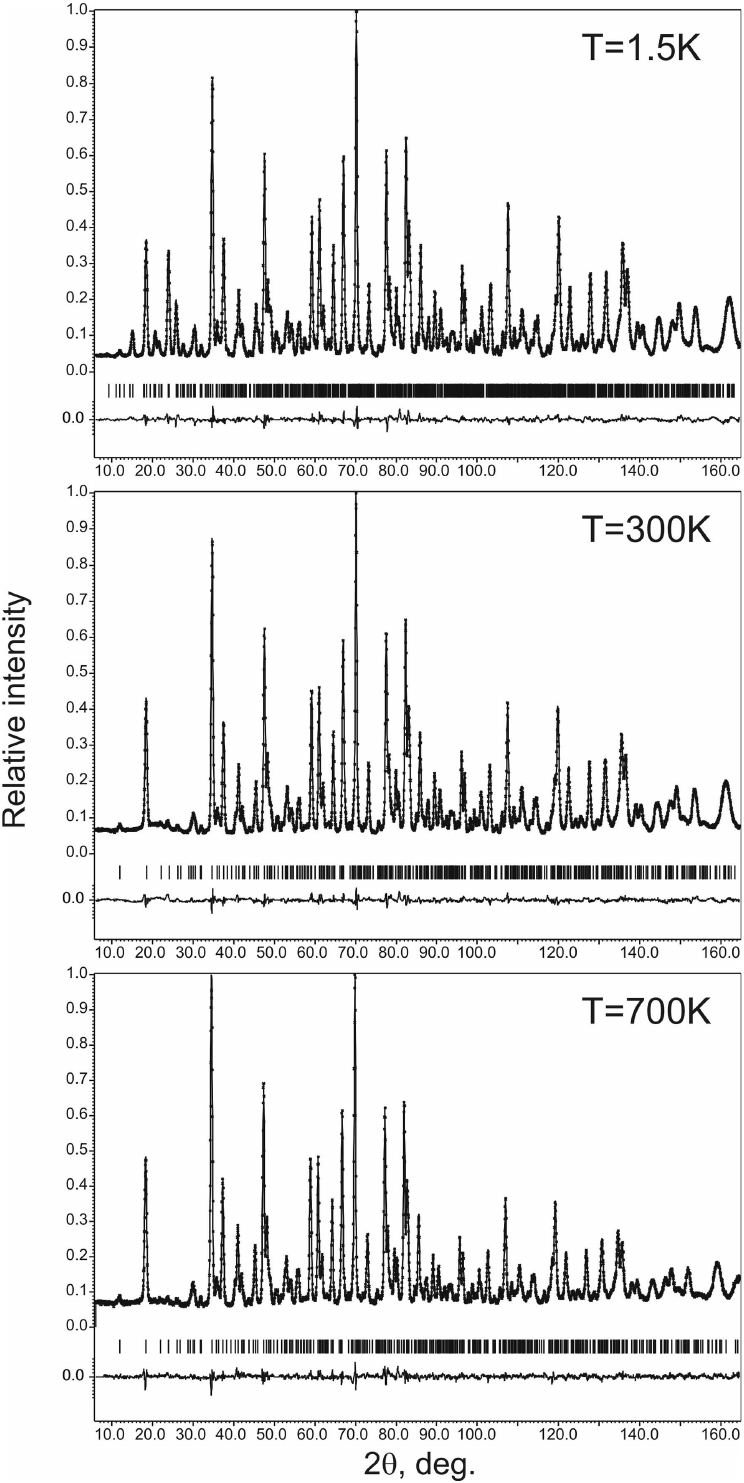}
\caption{\label{NPD}
Experimental, calculated and difference NPD patterns of Bi$_4$Fe$_5$O$_{13}$F after the Rietveld refinement at $T=1.5$~K, 300~K, and 700~K. The bars mark the reflection positions. For $T=1.5$~K the reflection positions of the magnetic supercell are shown.
}
\end{figure}

A preliminary XRPD investigation of the Bi$_4$Fe$_5$O$_{13}$F sample reveals a single-phase material with a tetragonal unit cell and the lattice parameters $a=8.30318(5)$~\r A, $c=18.0640(1)$~\r A. The electron diffraction patterns of Bi$_4$Fe$_5$O$_{13}$F are presented in Fig.~\ref{electron diffraction}. They can be consistently indexed on the tetragonal lattice with the unit cell parameters as determined from the XRPD data. The reflection conditions $0kl$, $k=2n$ and $hhl$, $l=2n$ suggest two possible space groups, $P4_2/mbc$ and $P4_2bc$. The structure solution was performed in the centrosymmetric $P4_2/mbc$ space group with direct methods providing the positions of all cations and part of the anions in the unit cell. The missing anions were located using difference Fourier maps. The resulting structural model was refined against the SXPD data at the temperatures of 10~K, 298~K and 673~K~\cite{supplemental}  and against the NPD data at the temperatures of 1.5~K, 300~K and 700~K (Fig.~\ref{NPD}). A magnetic contribution was taken into account for the refinement of the crystal structure from the NPD data collected at $T=1.5$~K (see details in section III-B). Oxygen and fluorine anions can not be distinguished reliably from either X-ray or neutron powder diffraction. Therefore, the assignment of the fluorine position at $4b$ (0,0,$\frac14$) is based on bond-valence-sum (BVS) arguments.   

BVS arguments strongly favor a complete ordering of the O and F anions. The F$^-$ anions linked to four Bi cations acquire a BVS of 0.91(2), which is close to their nominal valence. For the O$^{2-}$ anions occupying this position, the BVS would be as low as 1.20(2), thus demonstrating a strong and unrealistic underbonding. The anion ordering is additionally supported by the room-temperature M\"{o}ssbauer spectrum that is consistent with only two distinct coordination types of Fe$^{3+}$ cations (see below).

The refinement of the low-temperature diffraction data resulted in low atomic displacements parameters (ADPs). However, at and above room temperature high ADPs of F1 and O1 indicated possible displacements of these atoms. Indeed, the displacements of F1 and O1 from their special positions reduced the ADPs. This effect is mostly of dynamic nature, because at low temperatures the displacements are suppressed. The O1 and F1 displacements observed at high temperatures may be driven by subtle adjustments of interatomic distances and coordination (see BVS analysis in Supplemental Material~\cite{supplemental}), although further studies of the local structure would be necessary to fully elucidate this issue. We have also checked the possibility of symmetry lowering. The refinement in the acentric group $P4_2bc$ neither eliminated the disorder, nor reduced the reliability factors. Electron diffraction patterns clearly demonstrate the presence of glide planes perpendicular to the $a$ and $b$ unit cell axes, as well as perpendicular to the diagonals of the $ab$ face. Therefore, any further symmetry reduction can be safely excluded.

The crystallographic information for Bi$_4$Fe$_5$O$_{13}$F is summarized in Table~\ref{rietveld parameters}. Atomic parameters and main interatomic distances are provided in Tables~\ref{atomic parameters} and ~\ref{distances}. Further details on the structure refinement are given in Supplemental Material.~\cite{supplemental}

\begingroup
\begin{table*}
\begin{minipage}{15cm}
\caption{\label{atomic parameters}
Atomic positions and displacement parameters ($U_{\iso}$) of Bi$_4$Fe$_5$O$_{13}$F at 1.5~K, 300~K and 700~K.}
\begin{ruledtabular}
\begin{tabular}{cccc}
	$T$ (K) & 1.5 & 300 & 700 \\\hline
	Bi1, 16$i$ $(x,y,z)$ & & & \\
  $x/a$ & 0.6693(1) & 0.6706(2) & 0.6713(2) \\
  $y/b$ & 0.6594(1) & 0.6592(2) & 0.6592(2) \\
  $z/c$ & 0.15718(6) & 0.15697(7) & 0.15676(8) \\\medskip
  $U_{\iso}$ (\r A$^2$) & 0.001 & 0.0064(3) & 0.0195(3) \\
  Fe1, 8$f$ $(\frac12,0,z)$ & & & \\
  $z/c$ & 0.08014(9) & 0.0800(1) & 0.0798(1) \\\medskip
  $U_{\iso}$ (\r A$^2$) & 0.001 & 0.0049(4) & 0.0173(5) \\
  Fe2, 8$h$ $(x,y,0)$ & & & \\
  $x/a$ & 0.8511(2) & 0.8515(2) & 0.8516(3) \\
  $y/b$ & 0.8385(2) & 0.8388(2) & 0.8390(3) \\\medskip
  $U_{\iso}$~(\r A$^2$) & 0.001 & 0.0058(4) & 0.0201(5) \\
  O1 & $4a$ $(0,0,0)$ & $8e^*$ $(0,0,z)$ & $8e^*$ $(0,0,z)$ \\
  & & $z/c=0.0072(6)$ & $z/c=0.0106(6)$ \\\medskip
  $U_{\iso}$ (\r A$^2$) & 0.001 & 0.014(1) & 0.035(2) \\
  O2, $16i$ $(x,y,z)$ & & & \\
  $x/a$ & 0.2918(2) & 0.2914(2) & 0.2896(3) \\
  $y/b$ & 0.8764(2) & 0.8760(3) & 0.8761(3) \\
  $z/c$ & 0.5865(1) & 0.5858(1) & 0.5859(2) \\\medskip
  $U_{\iso}$~(\r A$^2$) & 0.001 & 0.0092(5) & 0.0225(7) \\
  O3, $8h,x,y,0$ & & & \\
  $x/a$ & 0.1347(3) & 0.1362(3) & 0.1368(4) \\
  $y/b$ & 0.5907(3) & 0.5900(3) & 0.5899(4) \\\medskip
  $U_{\iso}$ (\r A$^2$) & 0.001 & 0.0065(6) & 0.0189(8) \\
  O4, $8g$ $(x,x+\frac12,\frac34)$ & & & \\
  $x/a$ & 0.3263(3) & 0.3268(3) & 0.3271(3) \\\medskip
  $U_{\iso}$ (\r A$^2$) & 0.001 & 0.0077(6) & 0.0242(9) \\
  O5, $16i$ $(x,y,z)$ & & & \\
  $x/a$ & 0.5814(2) & 0.5823(2) & 0.5827(3) \\
  $y/b$ & 0.8602(2) & 0.8603(2) & 0.8592(3) \\
  $z/c$ & 0.3331(1) & 0.3331(2) & 0.3331(2) \\\medskip
  $U_{\iso}$~(\r A$^2$) & 0.001 & 0.0076(4) & 0.0207(5) \\
  F1 & $8e^*$ $(0,0,z)$ & $16i^{**}$ $(x,0,z)$ & $16i^{**}$ $(x,0,z)$ \\
  & $z/c=0.2587(3)$ & $x/a=0.024(3)$ & $x/a=0.022(6)$ \\
  & & $z/c=0.2597(7)$ & $z/c=0.260(1)$ \\
  $U_{\iso}$~(\r A$^2$) & 0.001 & 0.007(5) & 0.04(1) \\
\end{tabular}
\end{ruledtabular}
\flushleft{Reduced occupancy factors:$\quad$ $^*g$(F1)$=\frac12$ $\quad$ $^{**}g$(F1)$=\frac14$ $\quad$ $^{***}g$(O1)$=\frac12$}
\end{minipage}
\end{table*}
\endgroup

\begin{table}
\caption{\label{distances}
Selected interatomic distances (in~\r A) for Bi$_4$Fe$_5$O$_{13}$F at 1.5~K, 300~K, and 700~K.}
\begin{ruledtabular}
\begin{tabular}{cccc}
 $T$ (K) & 1.5 & 300 & 700 \\\hline
 Bi1--O2 & 2.173(2) & 2.173(3) & 2.176(3) \\
 Bi1--O4 & 2.160(4) & 2.125(2) & 2.133(2) \\
 Bi1--O5 & 2.170(2) & 2.168(2) & 2.171(3) \\
 Bi1--O5 & 2.551(2) & 2.555(2) & 2.557(3) \\
 Bi1--F1 & 2.456(4) & 2.35(1) & 2.37(3) \\
 Bi1--F1 & 2.658(4) & 2.57(2) & 2.56(3) \\
 Bi1--F1 & -- & 2.58(1) & 2.61(3) \\
 Bi1--F1 & -- & 2.79(2) & 2.80(3) \\\smallskip
 & & BVS = 3.15(1) & \\
 Fe1--O2$\times$2 & 2.009(2) & 2.017(2) & 2.035(2) \\  
 Fe1--O3$\times$2 & 1.975(2) & 1.981(2) & 1.986(3) \\
 Fe1--O5$\times$2 & 2.064(2) & 2.067(3) & 2.084(4) \\\smallskip
 & & BVS = 2.964(8) & \\
 Fe2--O1 & 1.820(1) & 1.824(2) & 1.833(3) \\
 Fe2--O2$\times$2 & 1.912(2) & 1.900(2) & 1.900(3) \\
 Fe2--O3 & 1.887(3) & 1.882(3) & 1.884(4) \\\smallskip
 & & BVS = 2.93(1) & \\    
 Fe3--O4$\times$2 & 2.035(3) & 2.032(2) & 2.036(3) \\
 Fe3--O5$\times$2 & 2.010(2) & 2.015(2) & 2.028(3) \\
 & & BVS = 2.956(7) & \\    
\end{tabular}
\end{ruledtabular}
\end{table}

\begin{figure*}
\includegraphics[width=16cm]{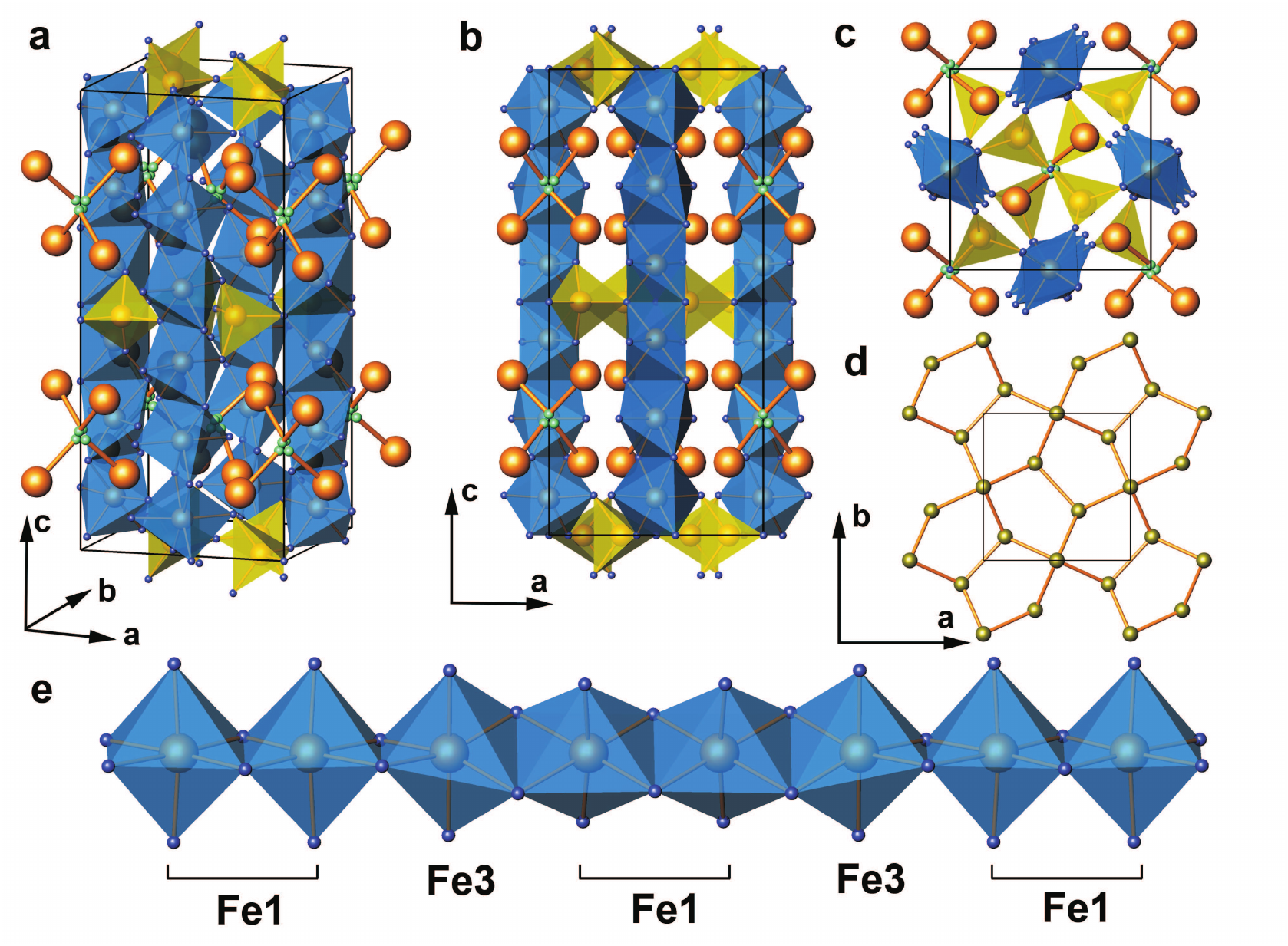}
\caption{\label{structure}
(Color online) The crystal structure of Bi$_4$Fe$_5$O$_{13}$F: clynographic viewing (a); [010] projection (b); [001] projection (c); Cairo pentagonal lattice of the Fe atoms (d) and enlarged view on the rutile-like chain (e). FeO$_6$ octahedra and FeO$_4$ tetrahedra are shown in blue and yellow, respectively. The Bi, F, Fe, and O atoms are drawn as large orange, small light green, medium green and small blue spheres, respectively.
}
\end{figure*}

The crystal structure of Bi$_4$Fe$_5$O$_{13}$F is illustrated in Fig.~\ref{structure}a-e. It is built of infinite rutile-like chains of edge-sharing FeO$_6$ octahedra running along the $c$-axis and comprising two Fe positions, Fe1 and Fe3 (Fig.~\ref{structure}e). The Fe2 atoms form pairs of corner-sharing FeO$_4$ tetrahedra that link the octahedral chains into a framework. These links repeat at every third FeO$_6$ octahedron of the chain, thus creating large cavities in the framework. The cavities are occupied by the fluorine-centered tetrahedral Bi$_4$F groups.

The structure of Bi$_4$Fe$_5$O$_{13}$F viewed along the $c$ direction immediately reveals the pentagonal arrangement of the magnetic Fe$^{3+}$ cations (compare Fig.~\ref{structure}d and Fig.~\ref{lattice}a). While in Bi$_2$Fe$_4$O$_9$ similar pentagonal planes are directly stacked on top of each other, Bi$_4$Fe$_5$O$_{13}$F demonstrates a more peculiar stacking, with the pentagonal planes interleaved by the Fe3O$_6$ octahedra. These octahedra disrupt the ideal eclipsed configuration within the rutile-like chains (Fig.~\ref{structure}c,e).

\begin{figure}
\includegraphics{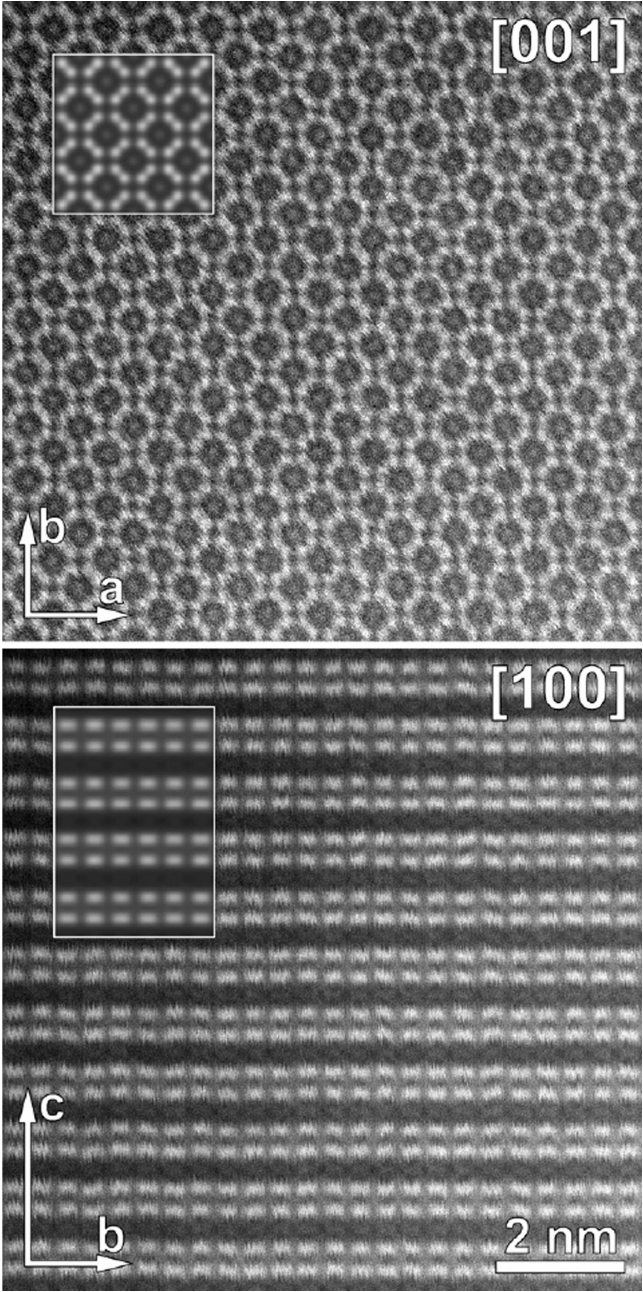}
\caption{\label{haadf}
[001] and [100] HAADF-STEM images of Bi$_4$Fe$_5$O$_{13}$F. The insets show the images calculated using the refined crystal structure. The size of the [001] calculated image is $3a\times3b$ (thickness $t$ = 4 nm); the size of the [100] calculated image is $3b\times2c$ (thickness $t$ = 7 nm).
}
\end{figure}

The Bi$_4$Fe$_5$O$_{13}$F crystal structure has been validated by HAADF-STEM imaging along the [001] and [100] directions (Fig.~\ref{haadf}). In HAADF-STEM, the brightness of a dot corresponding to the projection of an atomic column roughly scales as $Z^n$ ($1.5<n<2$), where $Z$ is the average atomic number along the column. Therefore, Bi-bearing atomic columns appear as the brighter dots and form the most prominent contrast in the image. The insets in Fig.~\ref{haadf} show HAADF-STEM images calculated using the refined atomic positions. The good agreement between the experimental and calculated images justifies the structure solution.

The room-temperature M\"{o}ssbauer spectrum for Bi$_4$Fe$_5$O$_{13}$F~\cite{supplemental} can be decomposed into two paramagnetic doublets, contributing to 57(4)~\% (doublet I) and 43(4)~\% (doublet II) of the total absorption area. M\"{o}ssbauer parameters, such as center shift (CS) and quadrupole splitting (QS), are as follows:

Doublet I: CS = 0.428(9) mm/s, QS = 0.39(2) mm/s,

Doublet II: CS = 0.19(1) mm/s, QS = 0.65(2) mm/s.

The CS values for both doublets evidence that all iron in the compound is in the Fe$^{3+}$ state.\cite{dickson2005,menil1985} The CS of the doublet I is consistent with Fe$^{3+}$ in an octahedral coordination.\cite{menil1985} The significantly smaller CS of the doublet II is in agreement with the tetrahedrally coordinated Fe$^{3+}$ (Ref.~\onlinecite{menil1985}). The 57(4):43(4) ratio of the octahedrally and tetrahedrally coordinated iron corresponds well to the anticipated 60:40 ratio in Bi$_4$Fe$_5$O$_{13}$F.

\subsection{Magnetic structure}

Below $T_N=178$~K, extra reflections appear on the NPD pattern. These reflections originate from a magnetic ordering, as no changes have been seen in the SXPD patterns down to $T=10$~K. The magnetic reflections in the entire temperature range of $T=1.5- 178$~K can be indexed with a $\mathbf{k}=(\frac12,\frac12,0)$ propagation vector. The magnetic structure at $T=1.5$~K was solved using a tetragonal magnetic supercell $\mathbf a_m=\mathbf a-\mathbf b$, $\mathbf b_m=\mathbf a+\mathbf b$, $\mathbf{c}_m=\mathbf c$. The analysis of possible magnetic symmetries was performed with the ISODISTORT software.\cite{campbell2006} The solution was found in the magnetic space group $P_C4_2/n$.~\cite{supplemental}  

The magnetic moments of the Fe positions, which are symmetrically equivalent in the nuclear structure, were kept identical. Although the symmetry does not require identical magnetic moments for Fe1 and Fe3, we found that these moments are equal within the standard deviation. Therefore, the corresponding linear constraint was introduced without affecting the fit quality. The refined magnetic moments at $T=1.5$~K are 4.06(6)~$\mu_B$ for the Fe1 and Fe3 positions and 3.34(6)~$\mu_B$ for the Fe2 position, which is significantly smaller than the expected moment of 5~$\mu_B$ for Fe$^{3+}$ in the high-spin configuration. Details on the refined magnetic moment components are listed in Table~\ref{magnetic moments}.

\begin{table}
\caption{\label{magnetic moments}
Positions of the magnetic atoms, the components of their magnetic moments, and total magnetic moments $(M)$ in Bi$_4$Fe$_5$O$_{13}$F at 1.5~K ($a = b = 11.7192$~\r A, $c = 18.0501$~\r A, all $m_z = 0$).}
\begin{ruledtabular}
\begin{tabular}{lcccrrr}
 Atom& $x/a$ & $y/b$ & $z/c$ & $m_x$ & $m_y$ & $M(\mu_B)$ \\\hline
 Fe1$_1$ & $\frac14$ & $\frac14$ & 0.0801 & 4.05(2) & $-0.35(6)$ & 4.06(6) \\
 Fe1$_2$ & $\frac14$ & $\frac14$ & 0.4199 & 4.05(2) & $-0.35(6)$ & 4.06(6) \\
 Fe2$_1$ & 0.0063 & 0.8448 & 0 & 2.18(4) & $-2.53(4)$ & 3.34(6) \\
 Fe2$_2$ & 0.0063 & $-0.3448$ & $\frac12$ & 2.53(4) & $-2.18(4)$ & 3.34(6) \\
 Fe3 & $\frac14$ & $\frac14$ & $\frac14$ & 4.05(2) & 0.35(6) & 4.06(6) \\
\end{tabular}
\end{ruledtabular}
\end{table}

\begin{figure}
\includegraphics{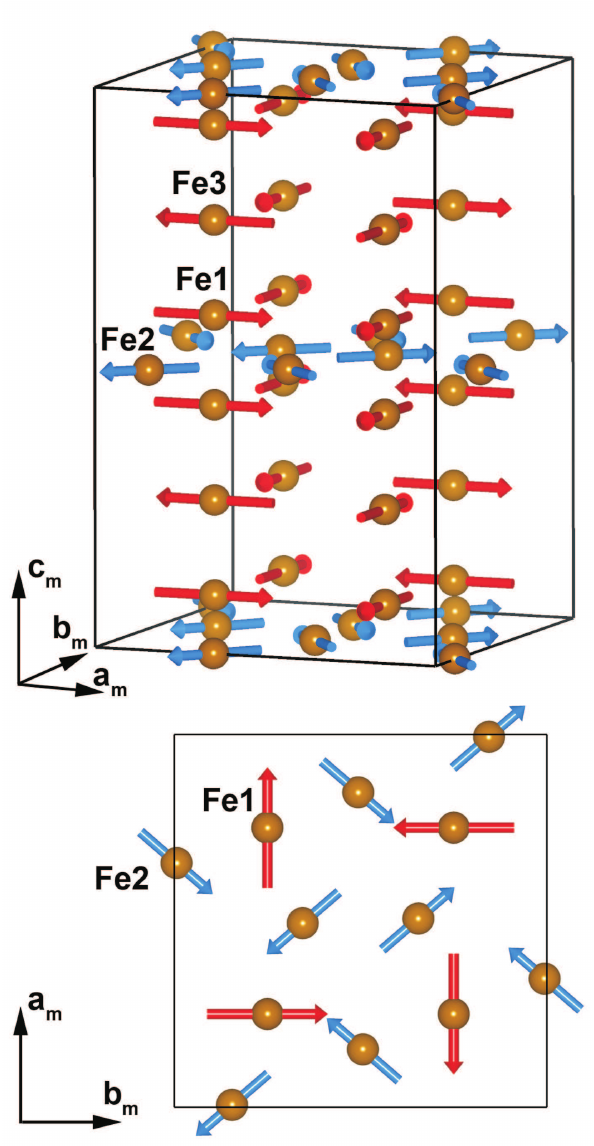}
\caption{\label{magnetic structure}
(Color online) The magnetic structure of Bi$_4$Fe$_5$O$_{13}$F at $T=1.5$~K. The orientation of the magnetic moments within the pentagonal Fe1--Fe2 layer at $z=\frac12$ is shown at the bottom.
}
\end{figure}

The peculiar non-collinear antiferromagnetic structure of Bi$_4$Fe$_5$O$_{13}$F is illustrated in Fig.~\ref{magnetic structure}. All magnetic moments are confined to the $ab$ plane. The order along the rutile-type chains is ferrimagnetic, with two parallel magnetic moments on the neighboring Fe1 atoms, and an opposite moment on Fe3. The direction of these moments is close to [110] and rotates by 90$^{\circ}$ between neighboring chains, so that the moments in second-neighbor chains are opposite and cancel each other. The magnetic moments of the Fe2 atoms are $\sim4.2^{\circ}$ off the {100} direction and form an AFM configuration within each Fe$_2$O$_7$ dimer (Fig.~\ref{magnetic structure}).

\begin{figure}
\includegraphics{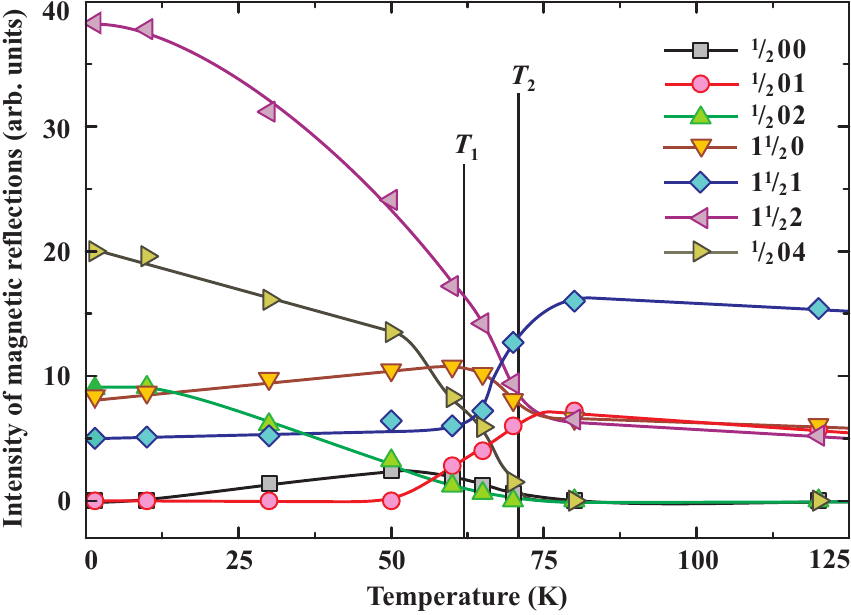}
\caption{\label{magnetic reflections}
(Color online) Temperature variation of the intensity of the magnetic reflections.
}
\end{figure}

The temperature evolution of the magnetic reflections is non-monotonic and pinpoints at least two additional magnetic transitions that follow the initial AFM ordering at $T_N=178$~K (Fig.~\ref{magnetic reflections}). The transition at $T_1\approx 62$~K is manifested by abrupt intensity variations: the reflections with even $l$ rapidly vanish, whereas the intensities of the reflections with odd $l$ notably increase. The second transition at $T_2\approx 71$~K manifests itself by the vanishing of the $\frac 12\,0\,l,\, l=2n$ reflections. The two magnetic transitions below $T_N$ are well in line with the results of thermodynamic measurements (Section~\ref{sec:thermo}). These transitions are likely accompanied by drastic changes in the magnetic structure. Details of this complex and intriguing magnetic behavior should be addressed in future studies.

\subsection{Thermodynamic properties}
\label{sec:thermo}
\begin{figure}
\includegraphics{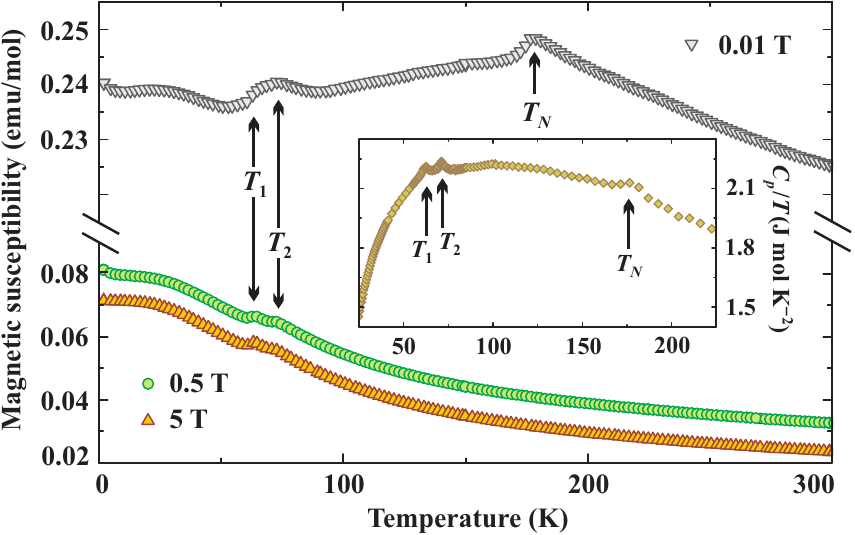}
\caption{\label{magnetic measurements}
(Color online) Magnetic susceptibility of Bi$_4$Fe$_5$O$_{13}$F measured in applied fields of 0.01~T, 0.5~T, and 5~T. The sizable field dependence is due to a tiny amount of ferromagnetic impurity. The inset shows the heat capacity divided by temperature ($C_p/$T). Arrows denote the transition temperatures $T_1$, $T_2$, and $T_N$ (compare to Fig.~\ref{magnetic reflections}).
}
\end{figure}

\begin{figure}
\includegraphics{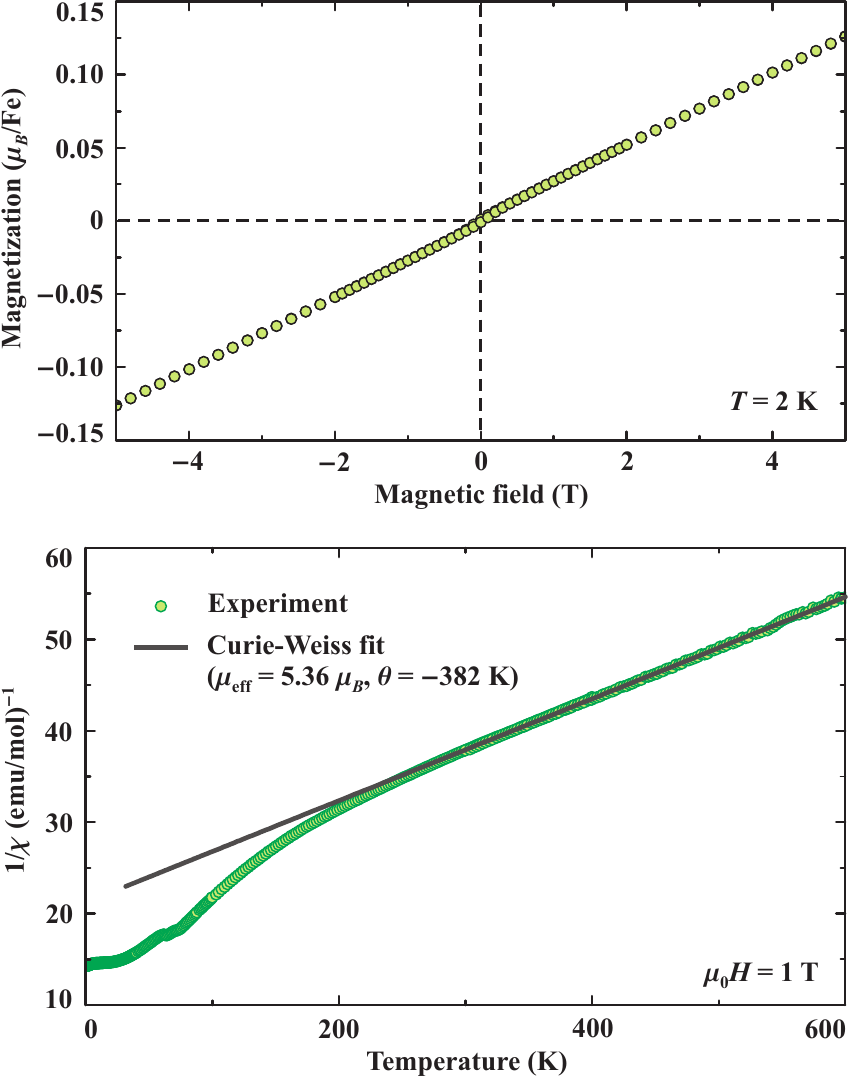}
\caption{\label{magnetic measurements-II}
(Color online) Magnetization curve of Bi$_4$Fe$_5$O$_{13}$F measured at 2 K (top) and inverse magnetic susceptibility (bottom) measured in the applied magnetic field of 1~T (circles) along with the Curie-Weiss fit (solid line).
}
\end{figure}

The magnetic susceptibility of Bi$_4$Fe$_5$O$_{13}$F reveals an overall antiferromagnetic behavior (Fig.~\ref{magnetic measurements}) in agreement with the antiferromagnetic, albeit non-collinear spin arrangement, pinpointed by neutron diffraction. The field dependence of the susceptibility should be ascribed to a tiny amount of a ferromagnetic impurity, such as Fe$_2$O$_3$. Magnetization loops measured at constant temperature do not show any clear signature of a remnant magnetization or hysteresis (Fig.~\ref{magnetic measurements-II}, top).

Above 300~K, the susceptibility follows the Curie-Weiss law $\chi= C/(T+\theta)$ with $C$ = 3.59(3)~emu~K/mol Fe and $\theta=-380(10)$~K (Fig.~\ref{magnetic measurements-II}, bottom). The resulting Curie constant corresponds to an effective magnetic moment of $\mu_{\eff}=5.4$~$\mu_B$, in reasonable agreement with 5.92~$\mu_B$ expected for the spin-5/2 Fe$^{3+}$ cations. The negative $\theta$ indicates predominant AFM interactions in Bi$_4$Fe$_5$O$_{13}$F.

In fields of 0.1~T and higher (Fig.~\ref{magnetic measurements}, bottom part), the most conspicuous effect is the double anomaly at $T_1=62$~K and $T_2=71$~K. In lower fields, a third anomaly around $T_N=178$~K could be additionally observed (Fig.~\ref{magnetic measurements}, top). Heat-capacity ($C_p$) measurements consistently show the anomalies at $T_1$, $T_2$, and $T_N$, thus confirming the intrinsic nature of these magnetic transitions (Fig.~\ref{magnetic measurements}, inset). The anomalies at $T_1$ and $T_2$ are indicative of first-order transitions with small latent heats.

Considering the neutron diffraction data, we conclude that Bi$_4$Fe$_5$O$_{13}$F undergoes an antiferromagnetic ordering of the Fe moments at $T_N=178$~K. However, this ordering is presumably incomplete, as shown by the large anomalies in the heat capacity at $T_1$ and $T_2$. The area under the heat-capacity curve plotted as $C_p/T$ vs. $T$ corresponds to the entropy release upon a magnetic transition. The large amount of entropy released at $T_1$ and $T_2$ implies that above $T_1$ the system develops an incompletely ordered intermediate AFM structure, which is further altered at $T_2$ and eventually disappears at $T_N$. This peculiar behavior is independently confirmed by the drastic changes in the magnetic neutron scattering (Sec.III-B). 

\subsection{Microscopic magnetic model}

\begin{figure}
\includegraphics{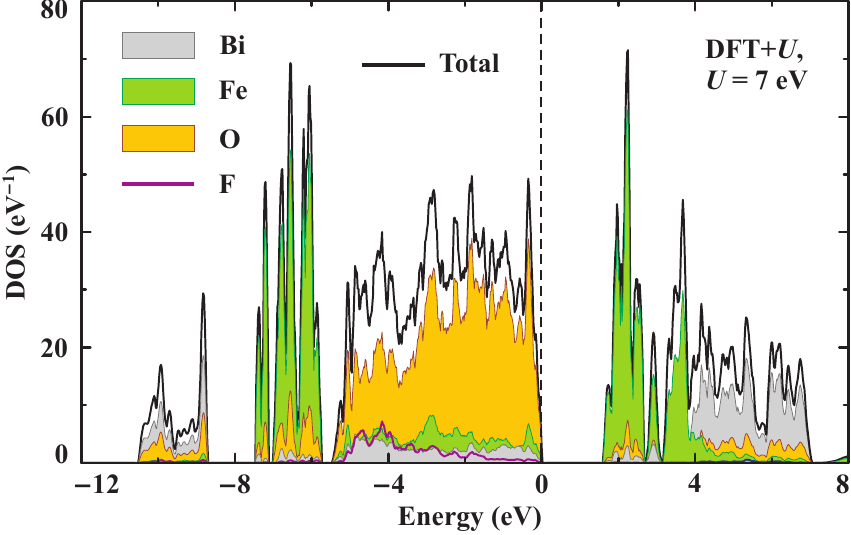}
\caption{\label{DOS}
(Color online) Electronic density of states calculated for the lowest-energy collinear AFM spin configuration of Bi$_4$Fe$_5$O$_{13}$F with DFT+$U$ ($U = 7$~eV). The Fermi level is at zero energy. Only one spin direction is shown.
}
\end{figure}

To elucidate relevant magnetic interactions, the band structure of Bi$_4$Fe$_5$O$_{13}$F was calculated. Regarding the band structure, this oxyfluoride is very similar to insulating Fe$^{3+}$ oxides. Its electronic spectrum (Fig.~\ref{DOS}) reveals oxygen and fluorine $2p$ bands between $-5$~eV and the Fermi level. The Fe $3d$ bands are around $-6$~eV (filled) and at $+2$~eV (empty), with the splitting of 8~eV that roughly matches the on-site Coulomb repulsion of $U=7$~eV. Bi $6s$ orbitals form bonding states at around $-10$~eV. In DFT+$U$ calculations, the energy gap of $1.5-2.0$~eV depending on the spin configuration concurs with the brown color of Bi$_4$Fe$_5$O$_{13}$F.

Magnetic couplings extracted from the DFT+$U$ band structure indicate a predominant interaction $J_d$ = 252~K within the Fe$_2$O$_7$ dimer, in agreement with the 180$^{\circ}$ Fe--O--Fe exchange pathway, which is most favorable for the AFM coupling, according to the Goodenough-Kanamori-Anderson rules. Other nearest-neighbor couplings are also AFM, but much weaker than $J_d$ according to the lower Fe--O--Fe angles (see Table~\ref{integrals}). In general, the couplings decrease as the Fe--O--Fe angle approaches 90$^{\circ}$, although this trend is violated at the lowest angles ($J_{c1}>J_{c2}$), where the direct Fe-Fe exchange comes into play. The couplings beyond nearest neighbors are all below 2~K and can be safely neglected in a first approximation. The resulting spin lattice has pentagonal-like units and 3 nonequivalent couplings in the $ab$ plane (Fig.~\ref{lattice}b). The interactions $J_{c2}$ connect these planes along the $c$ direction via the Fe3 spins, which do not belong to the pentagonal units.

\begin{table}
\caption{\label{integrals}
Magnetic interactions in Bi$_4$Fe$_5$O$_{13}$F. The Table lists Fe--Fe distances, Fe--O--Fe angles, exchange integrals $J_i$, as well as normalized spin-spin correlations $\langle S_iS_j\rangle/S^2$ in the ground-state magnetic configuration, as obtained from Monte-Carlo simulations.}
\begin{ruledtabular}
\begin{tabular}{ccccc}
 Interaction & $d_{\text{Fe--Fe}}$ (\r A) &  $\varphi_{\text{Fe-O-Fe}}$ (deg) & $J_i$ (K) & $\langle S_iS_j\rangle/S^2$  \\\hline
 $J_{c1}$ & 2.91 & 94.2 & 34 & +0.99 \\
 $J_{c2}$ & 3.06 & 97.4 & 10 & $-0.94$ \\
 $J_{ab1}$ & 3.39 & 119.2 & 45 & $-0.84$ \\
 $J_{ab2}$ & 3.53 & 130.9 & 74 & $-0.80$ \\
 $J_d$ & 3.64 & 180 & 191 & $-0.97$ \\
\end{tabular}
\end{ruledtabular}
\end{table}

Monte-Carlo simulations for the spin lattice depicted in Fig.~\ref{lattice}b and the $J_i$ parameters from Table~\ref{integrals} suggest that at low temperatures Bi$_4$Fe$_5$O$_{13}$F should be magnetically ordered. The ordering pattern can be extracted from the spin-spin correlations (Table~\ref{integrals}), as explained in Ref.~\onlinecite{abakumovl2011}. The normalized correlation $\langle S_iS_j\rangle/S^2$ is the measure of $\cos\varphi$, where $\varphi$ is the angle between the magnetic moments. Negative normalized correlations approaching $-1$ indicate antiparallel magnetic moments within the Fe$_2$O$_7$ dimers (the interaction $J_d$) as well as between Fe1 and Fe3 along the $c$ direction (the interaction $J_{c2}$). Despite the sizable AFM coupling $J_{c1}$ between the adjacent Fe1 sites along $c$, the moments on these Fe1 sites are parallel, as shown by the positive spin-spin correlation of nearly +1 (Table~\ref{integrals}). In this way, the simulations identify the ferrimagnetic order within the rutile-like chains, in agreement with the experiment.

Following the expectations, the spin lattice of Bi$_4$Fe$_5$O$_{13}$F features pentagonal units with exclusively AFM interactions (Fig.~\ref{lattice}b) that lead to a strong magnetic frustration. Indeed, in the ground-state magnetic structure only the strongest coupling $J_d$ is fully satisfied. The moments coupled by $J_{ab1}$ and $J_{ab2}$ develop the non-collinear configuration, whereas the AFM coupling $J_{c1}$ is fully suppressed and gives way to the FM order. The coupling $J_{c2}$ is also fully satisfied, because it does not compete with any other interaction in this system. 

\section{Discussion and summary}
\label{discussion}

Our experimental and theoretical results establish the strong magnetic frustration in Bi$_4$Fe$_5$O$_{13}$F. This effect  originates from the pentagonal units of the spin lattice. The frustration underlies the intricate non-collinear magnetic structure and the overall peculiar magnetic behavior with the magnetic ordering at $T_N$ and further transformations at $T_1$ and $T_2$.

Similar to Bi$_2$Fe$_4$O$_9$, Bi$_4$Fe$_5$O$_{13}$F is not a perfect prototype of the Cairo spin lattice (Fig.~\ref{lattice}). In Bi$_4$Fe$_5$O$_{13}$F, two nonequivalent couplings $J_{ab1}$ and $J_{ab2}$ stand for the single coupling $J_2$. Additionally, the lattice is decorated by Fe1 dumbbells that replace all four-vertex sites. However, this kind of decoration does not change any basic features of the magnetic system. The spin lattices of Bi$_2$Fe$_4$O$_9$ and Bi$_4$Fe$_5$O$_{13}$F are topologically equivalent to the Cairo lattice. The non-collinear magnetic structures observed in these compounds follow theoretical predictions for the ideal Cairo lattice.\cite{ressouche2009}

Two striking differences between Bi$_2$Fe$_4$O$_9$ and Bi$_4$Fe$_5$O$_{13}$F should be emphasized, though. Experimentally, Bi$_2$Fe$_4$O$_9$ orders antiferromagnetically below 264~K and retains the same magnetic structure down to the lowest temperatures. This can be seen from the smooth temperature evolution of the magnetic reflections.\cite{ressouche2009,shamir1978} In Bi$_4$Fe$_5$O$_{13}$F, the AFM ordering that sets in below $T_N=178$~K is likely incomplete and changes upon further magnetic transitions at $T_1$ and $T_2$. As none of these transitions is accompanied by any structural transformations detectable by NPD and SXPD, the sizable release of entropy and the indications for a weak first-order contribution imply that above $T_1$ the magnetic order in Bi$_4$Fe$_5$O$_{13}$F is altered. One plausible explanation is the formation of partially disordered AFM states that have higher entropy than the fully ordered non-collinear AFM structure. The partially disordered states are found in magnetic models based on triangular spin lattices.\cite{mekata1977,kaburagi1982,wada1983} They are indeed observed in Co$^{2+}$ halides, such as CsCoCl$_3$\cite{mekata1978} and RbCoBr$_3$.\cite{nishiwaki2008} Surprisingly, our experimental results may put forward similar physics on the pentagonal lattice. Our results also rise further questions to theory, because the sequence of magnetic transitions below the N\'eel temperature was never anticipated for any pentagonal geometry.

\begin{figure}
\includegraphics{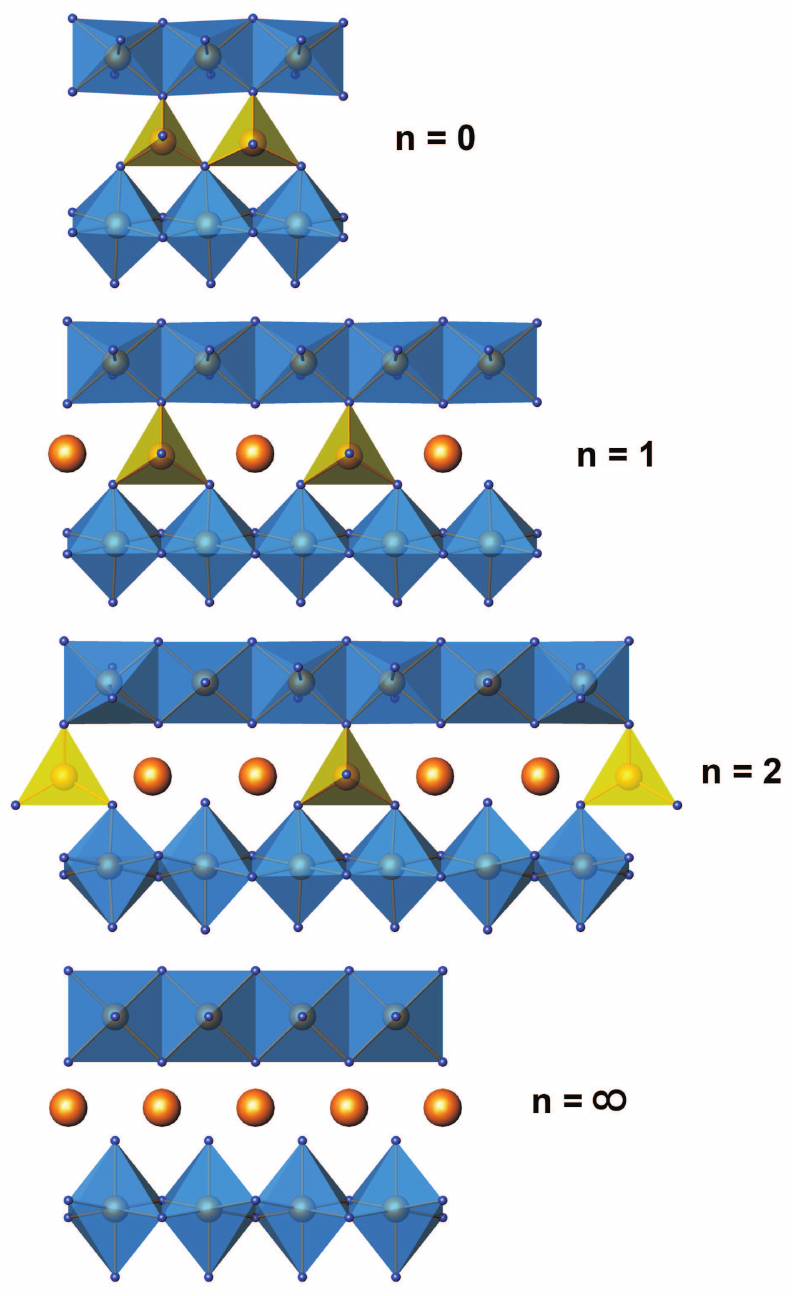}
\caption{\label{homologous}
(Color online) The scheme of the linkage between rutile-type octahedral chains (blue) through the tetrahedral units (yellow) in the B$_3$O$_5$ + $n$A$_2$BO$_4$ homologous series for different $n$. The A cations are shown as orange spheres.
}
\end{figure}

The second difference between Bi$_4$Fe$_5$O$_{13}$F and Bi$_2$Fe$_4$O$_9$ pertains to the stacking of the pentagonal layers. While in Bi$_2$Fe$_4$O$_9$ these layers are directly stacked on top of each other, Bi$_4$Fe$_5$O$_{13}$F features additional Fe3 sites between the pentagonal layers. The formation of these additional sites accompanies the incorporation of the Bi$_4$F tetrahedra. Indeed, a comparison between the crystal structures of Bi$_2$Fe$_4$O$_9$, Bi$_4$Fe$_5$O$_{13}$F, and sillimanite Al$_2$SiO$_5$ (the aristotype structure for the mullite-type compounds\cite{schneider2008}) opens broad perspectives for finding new systems with a pentagonal arrangement of magnetic atoms. All these structures contain infinite rutile-type chains of octahedra and differ by the connectivity of the chains through the B$_2$O$_7$ tetrahedral groups. In the Al$_2$SiO$_5$ (= B$_3$O$_5$) sillimanite structure, the connectivity is the densest: every BO$_6$ octahedron shares a corner with the tetrahedral group (Fig.~\ref{homologous}). In the Bi$_2$Fe$_4$O$_9$ structure, the linkages are repeated at every second octahedron, whereas in Bi$_4$Fe$_5$O$_{13}$F every third octahedron is linked.

The series of the Al$_2$SiO$_5$, Bi$_2$Fe$_4$O$_9$, and Bi$_4$Fe$_5$O$_{13}$F compounds can be formally considered as an insertion of A$_2$BO$_4$ fragments between the tetrahedral linkages. The A$_2$BO$_4$ entity includes two atomic layers, A$_2$O$_2$ and BO$_2$, and increases the distance between the tetrahedral linkages. Additionally, two A-cations located between the chains appear. This way, one arrives at the sillimanite-based homologous series B$_3$O$_5$ + $n$A$_2$BO$_4$, where $n = 0$ corresponds to the sillimanite structure, $n = 1$ comprises a group of Bi$_2$Fe$_4$O$_9$ (B = Fe, Al, Ga) compounds\cite{tutov1970, niizeki1968, arpe1977, abrahams1999, muller1978} and $n$ = 2 is represented by Bi$_4$Fe$_5$O$_{13}$(F), where the extra F atom is added between two A$_2$O$_2$ layers. Extending this series to $n = \infty$, one arrives at the A$_2$BO$_4$ structure, where tetrahedral linkages between the rutile-type chains are absent. This structure is exemplified by Sb$_2$BO$_4$ (B = Mn, Fe, Ni)\cite{chater1985, fjellvag1985, chater1985a, gavarri1982} belonging to the Pb$_3$O$_4$ structure type.

The above crystallographic arguments suggest that other members of the homologous series should be formed in between the limiting cases of B$_3$O$_5$ ($n = 0$) and A$_2$BO$_4$ ($n = \infty$). Indeed, the $n = 1$ (Bi$_2$Fe$_4$O$_9$) and $n = 2$ (Bi$_4$Fe$_5$O$_{13}$F) members are already available. The intermediate compounds would contain pentagonal-lattice units in the $ab$ plane connected via intermediate B sites along the $c$ direction. While Bi$_2$Fe$_4$O$_9$ shows a single magnetic transition, Bi$_4$Fe$_5$O$_{13}$F reveals a more intricate and peculiar physics with several transitions that can be retained or probably altered in further members of the family with higher $n$. This approach also enables the control over the dimensionality, because the separation between the pentagonal layers increases with $n$.

The compounds with high $n$ require appropriate A cations. Considering the available Bi$_2$Fe$_4$O$_9$ and Bi$_4$Fe$_5$O$_{13}$F compositions, the lone electron pair is likely a necessary prerequisite. Besides Bi$^{3+}$, Sb$^{3+}$ and Pb$^{2+}$ are plausible candidates because of their known compatibility with the A$_2$BO$_4$ structure.\cite{abakumov2005} The combination of Bi$^{3+}$, Sb$^{3+}$ and Pb$^{2+}$ enables heterovalent replacements and the fine tuning of the electron concentration. Another interesting feature is the ability to accommodate extra anions. The F$^-$ anions can be inserted into tetrahedral voids formed between two successive A$_2$O$_2$ layers, as in the $n = 2$ Bi$_4$Fe$_5$O$_{13}$F structure that has the A$_4$B$_5$O$_{13}$F composition instead of the expected A$_4$B$_5$O$_{13}$. Alternatively, the anion content can be changed by replacing the common vertex of two FeO$_4$ tetrahedra with an oxygen dumbbell that would transform a pair of corner-sharing tetrahedra into a pair of edge-sharing tetragonal pyramids B$_2$O$_8$. This replacement is observed in the Mn-based version of the $n = 1$ compound, Bi$_2$Mn$_4$O$_{10}$ (= A$_2$B$_4$O$_9$O).\cite{niizeki1968}

In summary, the discovery of Bi$_4$Fe$_5$O$_{13}$F not only provides an interesting material prototype of a frustrated Cairo pentagonal lattice, but also opens tantalizing opportunities for a further expansion of this interesting class of materials. At low temperatures, Bi$_4$Fe$_5$O$_{13}$F develops the non-collinear antiferromagnetic order, which is anticipated for the pentagonal Cairo lattice. However, this order is strongly altered at $T_1\approx 62$~K and $T_2\approx 71$~K. The respective magnetic transitions have no detectable structural component. They also involve the sizable increase in the magnetic entropy that may be ascribed to the transformation toward partially ordered antiferromagnetic states extending up to the N\'eel temperature of $T_N\approx 178$~K. Additionally, Bi$_4$Fe$_5$O$_{13}$F manifests the B$_3$O$_5$ + $n$A$_2$BO$_4$ homologous series of compounds that form further examples of pentagonal spin lattices, thus facilitating a precise tuning of the "dimensionality" as well as electronic concentration in these systems.

\acknowledgments

J.H. and D.B. acknowledge support from the Research Foundation -- Flanders, through project G.0184.09N. G.V.T. acknowledges the financial support from the ERC grant N$^{\circ}$246791-COUNTATOMS. A.T. was funded by Alexander von Humboldt Foundation and the Mobilitas grant of the ESF. A.T. is also grateful to Ioannis Rousochatzakis for drawing his attention to the physics of Cairo spin lattice. We are grateful to the ESRF and PSI for granting the beam-time. Experimental support of Andy Fitch at the ID31 beamline of ESRF is kindly acknowledged.

%

\end{document}